\documentclass[aps]{revtex4}
\usepackage{graphicx}
\usepackage[all]{xy}
\usepackage{amsmath}
\usepackage{amssymb}
\usepackage{color}

\newcommand{\bb}{\bibitem}
\newcommand{\bes}{\begin{subequations}}
\newcommand{\ees}{\end{subequations}}
\newcommand{\benn}{\begin{eqnarray*}}
\newcommand{\eenn}{\end{eqnarray*}}
\newcommand{\del}{\partial}
\def\ben{\begin{eqnarray}}
\def\een{\end{eqnarray}}
\def\be{\begin{equation}}
\def\ee{\end{equation}}

\begin{document}
\title{4D gravity on a BPS brane in 5D $AdS$-Minkowski space}
\author{R.C. Fonseca$^{a}$, F.A. Brito$^{b}$, and L. Losano$^a$}
\affiliation{{\small {{$^a$}{Departamento de F\'\i sica, Universidade Federal da Para\'\i ba, 58051-970 Jo\~ao Pessoa PB, Brazil}}\\
{$^b$}{Departamento de F\'\i sica, Universidade Federal de Campina
Grande, \\Caixa Postal 10071, 58109-970 Campina Grande, Para\'\i ba,
Brazil}}}

\begin{abstract}
We calculate small correction terms to gravitational potential near an asymmetric BPS brane embedded in a $5D$ $AdS$-Minkowski space in the context of supergravity. The normalizable wave functions of gravity fluctuations around the brane describe only massive modes. We compute such wave functions analytically in the thin wall limit. We estimate the correction to gravitational potential for small and long distances, and show that there is an intermediate range of distances in which we can identify $4D$ gravity on the brane below a crossover scale. The $4D$ gravity is metastable and for distances much larger than the crossover scale the $5D$ gravity is recovered.
\end{abstract}

\maketitle


Following the original idea of Randall and Sundrum (RS) \cite{RS1}, we consider a braneworld scenario. In the RS scenario the five-dimensional gravity is coupled to a negative cosmological constant and a 3-brane sourced by a delta function. The solution in such setup is a {\it symmetric} solution given in terms of two copies of $AdS_5$ spaces patched together along the 3-brane. Although in this setup the fifth dimension is infinite the volume of the 5D bulk space is finite because the geometry is warped. As a consequence this allows having graviton zero mode responsible for 4D gravity on the brane. This is not necessary true for spaces asymptotically flat, because no zero mode emerges anymore. This was first shown by Gregory-Rubakov-Sibiryakov (GRS) \cite{RG} and  Dvali-Gabadadze-Porrati (DGP) \cite{dvali}. The nice consequence of such an alternative setup is that 4D gravity on the 3-brane now emerges due to gravity massive modes and then is metastable. However, gravity massive modes can live long enough before escaping from the 3-brane to produce 4D gravity within a sufficient large scale --- the crossover scale.   

In the present study we investigate such a scenario in a consistently truncated 5D supergravity \cite{susy_domain,Ref1}, where the 3-brane appears as BPS solutions. They are solutions of first order equations that emerges through Killing spinor equations that preserves part of the supersymmetries and also satisfy Einstein equations. We shall focus on the bosonic sector with 5D gravity coupled to two real scalar fields \cite{bbg,027}. 

In our investigations we are mainly interested on induced 4D gravity on {\it asymmetric} brane solutions \cite{Cvetic:1996vr,Padilla:2004tp,CastilloFelisola:2004eg,Gabadadze:2006jm}. Such brane solutions have naturally appeared in the supergravity context in four- \cite{Cvetic:1996vr} and five-dimensions \cite{027} where the thick 3-brane is embedded in an asymptotically  five-dimensional $AdS$-Minkowski space. We shall consider the later case, because it allows the possibility of metastable 4D gravity as first pointed out in the GRS \cite{RG} and DGP \cite{dvali} scenarios. Because the five-dimensional space is asymptotically Minkowski on one side of the 3-brane its volume is infinite and then no gravity zero mode emerges. However, just as in GRS and DGP scenarios, we also have found 4D gravity that lives long enough within the crossover scale.

Thus, as emphasized in GRS and DGP scenarios, we shall focus on the main beautiful characteristic of the 4D metastable gravity, that is the fact that whereas gravity becomes four-dimensional for distances very much smaller than the crossover scale, it emerges as a five-dimensional gravity for distances very much larger than such scale. In doing so, we shall find the Newtonian potential induced by the gravity massive modes of a Schroedinger-like equation for the gravity fluctuations around the asymmetric 3-brane solution.

Le us consider the bosonic sector of the supergravity action for 
spacetimes in arbitrary $D$-dimensions $(D >3)$ coupled to ${\cal N}$ real scalar fields given by \cite{susy_domain,Ref1}
\be\label{s}
S=\int{d^Dx\;\sqrt{|g|}\left[\frac{1}{2\kappa^{D-2}}R-\frac12 g^{MN}\del_M\phi_i\del_N\phi_i -V(\phi_i)\right]},
\ee
where $\kappa = \frac1M_*$ is the $D$-dimensional Planck length, and the potential of the scalar fields are taken as
\be\label{v}
V(\phi_i)=2(D-2)^2\left[\left(\frac{\del W}{\del \phi_i}\right)^2-\kappa^{D-2}\left(\frac{D-1}{D-2}\right)W^2\right],
\ee
where $W(\phi_i)$ is the superpotential, and $\phi_i$, $i=1,2,..,{\cal N}$,  are the scalar fields.
We employ the generalized Randall-Sundrum metric:
\be\label{ref1}
ds^2_D = {\rm e}^{A(y)}\eta_{\mu\nu}dx^{\mu}dx^{\nu} + dy^2,
\ee
where ${\rm e}^{A(y)}$ is warp factor, $\mu,\, \nu = 0, 1, 2, ...,D - 2$ are indices on the $(D-2)$ -
brane. By using the action \eqref{s} with the metric \eqref{ref1}, 
we obtain the set of
equations
\bes
\be
\phi_i''+(D-1)A'\phi_i'-\frac{\del V}{\del\phi_i}=0,
\ee
\be
A''= - \frac{\kappa^{D-2}}{(D-2)}\,\phi'^2
\ee
and
\be
A'^2 = 2\,\frac{\kappa^{D-2}}{(D-1)(D-2)}\,\left(\frac12\,\phi'^2-V\right),
\ee
\ees
which are solved by the following first order equations obtained from the Killing spinor equations
\bes
\be\label{ref1.2}
\del_yA = \mp2\kappa^{D-2}W,
\ee 
\be\label{ref1.3}\del_y\phi_i = ±2(D - 2)\frac{\del W}{\del\phi_i},
\ee
\ees
assuming that the scalar fields only depend on the transverse coordinate $y$.
The graviton modes on $(D - 2)$ - branes are governed by a linearized
gravity equation of motion in arbitrary number of
dimensions $(D>3)$ given by \cite{susy_domain, Ref1}
\be\label{r1.3}
\del_M(\sqrt{-g}g^{MN}\del_N\Phi) = 0,
\ee where $\Phi$ describes the wave function of the graviton on
non-compact coordinates $ M,N =0, 1, 2, ...,D - 1$. 
Let us consider $\Phi=h(y)\varphi(x^{\mu})$ into (\ref{r1.3}) and the fact
that $\Box_{D-1}\varphi =m^2\varphi$, where $\Box_{D-1}$ is the flat Laplacian operator
on the tangent frame. Thus, the wave equation for
the graviton through the transverse coordinate $y$ reads
\be\label{ref1.4}
\frac{\del_y(\sqrt{-g}g^{yy}\del_y h(y))}
{\sqrt{-g}}
= -m^2|g^{00}|h(y).
\ee
This is our starting point to investigate both zero and
massive graviton modes on the branes.
Using the components of the metric (\ref{ref1}) into the equation
(\ref{ref1.4}) we have
\be\label{ref1.5}
\frac12(D-1)\del_yA\del_yh(y) + \del_yh(y) = -m^2\,{\rm e}^{-A(y)}h(y),
\ee
which, 
changing the metric (\ref{ref1}) by the conformally flat metric,
\be\label{ref1.6}
ds^2_D = {\rm e}^{\,A(z)}(\eta_{\mu\nu}dx^{\mu}dx^{\nu} + dz^2),
\ee
and employing the changes of variables:
$h(y)=\psi(z)\,{\rm e}^{-\frac{A(z)(D-2)}{4}}$ and $z(y)=
\int{
{\rm e}^{-\frac{A(y)}{2}}}dy$,
can be written as the Schroedinger-like equation,
\be\label{ref1.7}
- \del^2 _z\psi(z) + V(z)\psi(z) = m^2\psi(z),
\ee
where the potential $V(z)$ is given by
\be\label{ref1.8}
V(z) =
\frac{(D - 2)^2}{16}\,(\del_zA)^2 +\frac{D - 2}{4}\,\del^2_zA.\ee

\begin{figure}[!htb]
\begin{center}
\includegraphics[{height=4cm,width=6cm,angle=00}]{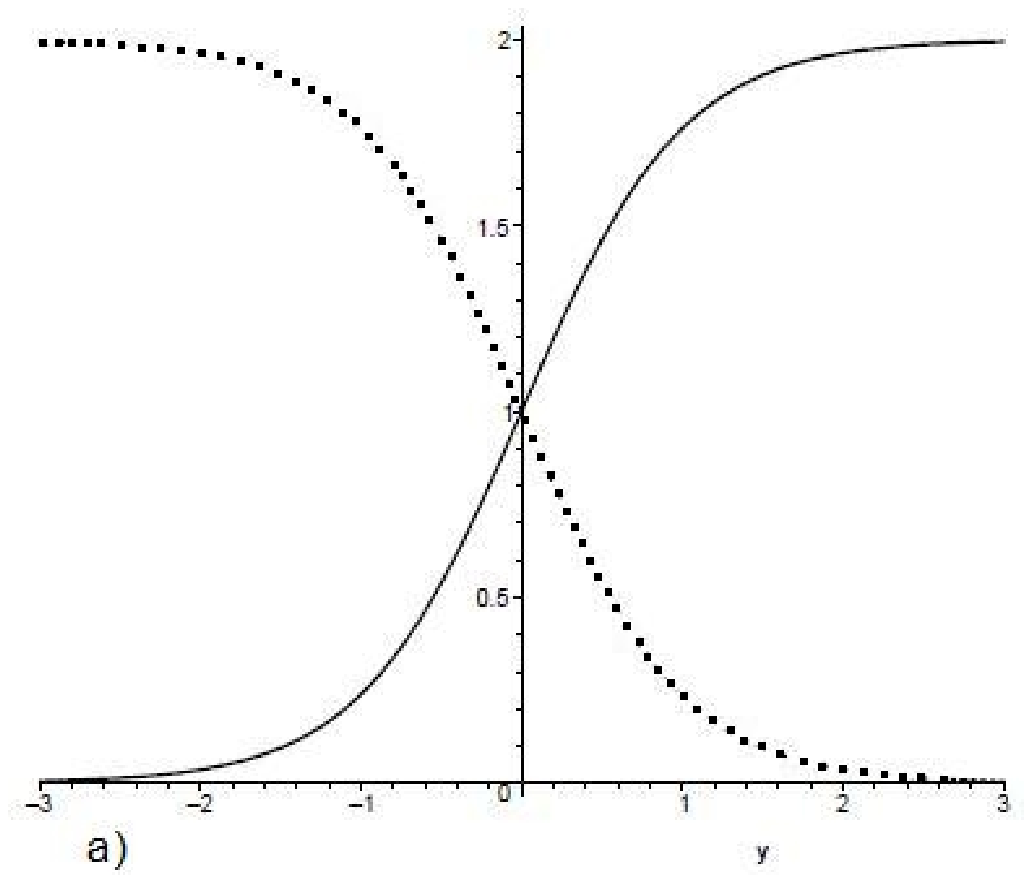}
\hspace{2.5cm}
\includegraphics[{height=4cm,width=6cm,angle=00}]{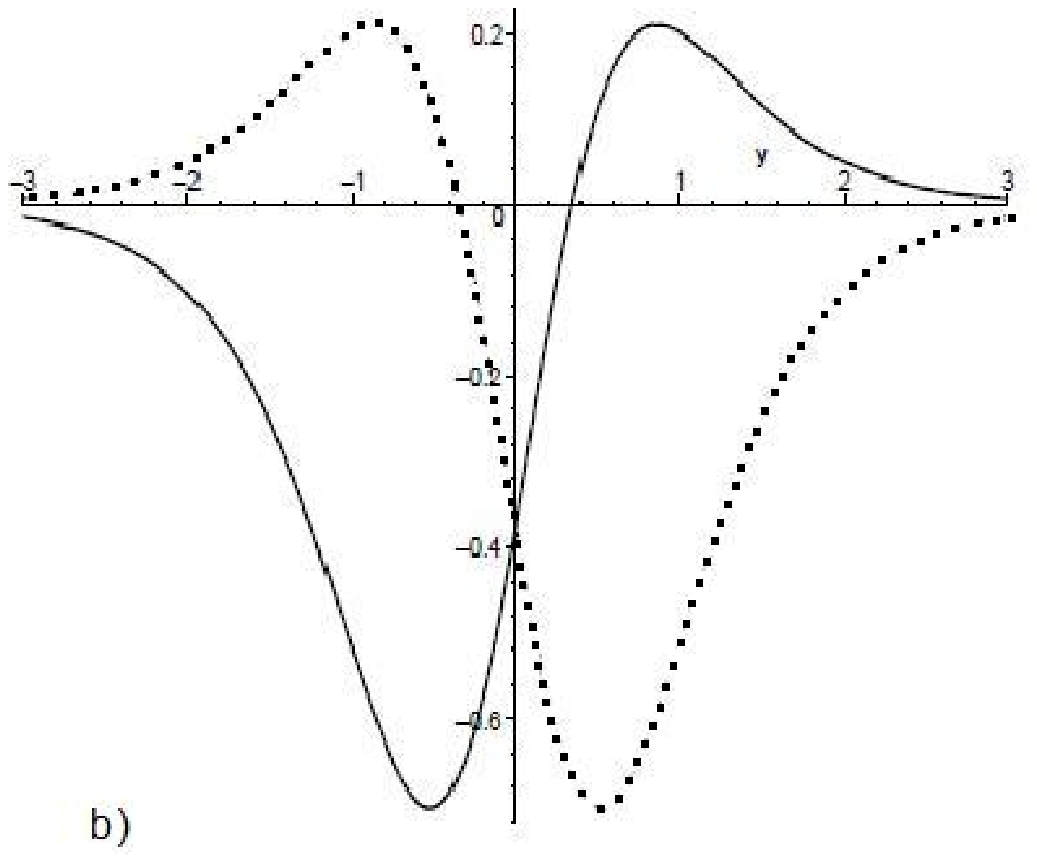}
\caption{\small a) Warp factor for BPS asymmetric branes connecting the spaces $AdS_5-\mathbb{M}_5$ (solid line)
and $\mathbb{M}_5-AdS_5$ (dashed line) asymptotically, with $a=1$ and
$b=2/\sqrt{3}$; b) Corresponding energy densities.}\label{fig3}
\end{center}
\end{figure}

Now, let us examine the theory introduced by the action \eqref{s}, for five-dimensional gravity coupled to two real scalar fields, $\phi_1$ and $\phi_2$. So let us make the transformation $W \rightarrow{W}/{[(D-1)(D-2)]}$, with $D=5$, in (\ref{ref1.2}) and (\ref{ref1.3}), and use units which $\kappa^{D-2}=2$. This is to be in accord with the model presented in
Ref.\cite{027} through the superpotential $W=3/2\,a\,\sin(\sqrt{2}\,b\,\phi_1)\cos(\sqrt{2}\,b\,\phi_2)$, where $a,b$ are real constants --- see also \cite{gonzalo} for other superpotentials.
The orbits $\cos(\sqrt{2}\,b\,\phi_1)=C\;\sin(\sqrt{2}\,b\,\phi_2)$, where $C$ is a real constant,  decouple  the first-order equations, and
for $C=1$ give the solutions 
\be\label{pq1}
\phi_1(y)=\pm\frac{\sqrt{2}}{4b}\arccos\left(\tanh\left(\frac34 ab^2y\right)\right)+(n+1)
\;\frac{\sqrt{2}\pi}{4b},\;\;\;\;\phi_2(y)=\pm\frac{\sqrt{2}}{4b}\arccos\left(\tanh\left(\frac32
ab^2y\right)\right)+n\;\frac{\sqrt{2}\pi}{4b}, \ee such that we obtain \be\label{ay5b}
A(y)=(-1)^{n+1}\frac{ay}{2}+\frac{2}{3b^2}\ln\left(\;q\;{\rm
sech}\left(\frac34ab^2y\right)\right), 
\ee where $q>0$. Then, we have BPS asymmetric branes connecting the spaces $AdS_5-\mathbb{M}_5$(Minkowski), see Fig.~\ref{fig3}, where the thin limit approximation, $\ln(\cosh(\varepsilon\,x))\rightarrow|\varepsilon\,x|$ for $\varepsilon>>1$, so that
\be\label{ap}
A(y)= \left( -1 \right) ^{n+1}\,\frac12\,a\,y+\frac23\,{\frac {\ln  \left( q \right) 
}{{b}^{2}}}-\frac12\, { 
 \left| a\,y \right| },\quad a\,b^2>>1,
\ee can be realized.
Here we have an additional constraint, $q\geq1$. We will continue the study considering a generic value of $q$ and then verify the threshold situations.
In order to investigate the existence of gravitational modes in the vicinity of the branes, assuming odd integer values for $n$, we employ a change of variable which makes the metric conformally flat by
\bes
\be\label{Z1}
y=\frac{2}{a\,k}\,z,\quad\quad z\leq0,
\ee
\be\label{Z2}
y=\frac2a\,\ln  \left( \frac{z}{k} +1
 \right), 
\quad\quad z>0.
\ee
\ees
The warp factor for each region, obtained by substituting the above equations in (\ref{ap}), taking $q=e^{3 ab^2k/4}$,
can be written in general form
\be\label{WP2}
{\it A} \left( z \right) =  -2\,\ln  \left( {
\frac{a\;(z+|z|)}{4}}+\frac{ ak}{2}\right).
\ee  
For $D=5$, replacing (\ref{WP2}) in (\ref{ref1.8}) and (\ref{ref1.7}), we obtain the Schroedinger-like equation for fluctuations given by
\be\label{POT1}
-\frac{d}{d\,z^2}\psi_m\left( z \right) +\left(\frac {15}{64}\,\frac{\left( 1+ {\rm sgn}(z)
  \right) ^{2}}{\left( \frac14\, \left( z+ \left| z \right| \right)+\frac{ k}{2}\right)^2}
-\frac38\,\frac{\delta \left( z \right) }{ \left( \frac14\, \left( z+ \left| z
 \right|  \right) +\frac{ k}{2}
 \right)}\right)\psi_m\left( z \right)=m^2\psi_m\left( z \right),
  \ee where $m^2$ is the four-dimensional mass which corresponds to the Kaluza-Klein modes. For $z<0$ the solution of the equation (\ref{POT1}), is given by
\bes  
\be\label{CONST}
\psi_{1_m} \left( z \right)=C_{{1}}\sin \left( m\,z \right) +C_{{2}}\cos \left( m\,z \right), 
\ee
and for $z>0$, we have the second-order Bessel equation whose solution is given by
\be\label{01}
\psi_{2_m} \left( z \right) =\sqrt{z+ k}\,\left[C_{{3}}\,J_{{2}} \left( m(z+ k)\right) +C_{{4}}\,N_{
{2}} \left( m(z+ k) \right)\right],
\ee
\ees
where $J_{2}$ and $N_{2}$ are Bessel functions of the first and second kind,
respectively, and $C_\ell$, $\ell=1,2,3,4$, are constants to be determined below. Starting from this point, we will investigate the particular solution in which $C_1=0$ which eliminates the term that does not contribute to the continuity condition at $z=0$ and also gives us $\psi_{1_m}'(0)=0$.  Moreover, as this is an oscillatory solution on the bulk, the phase on the brane can be chosen conveniently.
Since we are interested in the correction terms to the four-dimensional Newton law
between two unit masses on the brane, it is necessary to obtain the probability of
gravity with KK-modes on the brane. The asymptotic behavior of $|\psi_m\left(0\right)|^2$ depends on the magnitude of the argument in the Bessel functions and the
normalization factors $C_1$, $C_3$ and $C_4$ under certain conditions. The continuity condition of the wave function $\psi_{1_m}(0)=\psi_{2_m}(0)$ and the jump condition at $z=0$,
\be\label{jump}
\left[{\frac {d}{dz}}\psi_{2_m} \left( z \right)\right]_{{z=0}}+\frac34\,{\frac {\psi_{2_m}
 \left( 0 \right) }{ k}}=0,
\ee
lead to the relation between these constants. Furthermore, the normalization factors in wave function can also be determined by the
orthonormalization condition of Bessel functions. Then, after the above considerations we can rewrite (\ref{CONST}) and (\ref{01}) as
\bes\label{psi12}
\be\label{psi1}
\psi_{1_m} \left( z \right) =\frac{4}{\pi}\,{\frac {C_{{m}}}{\sqrt{ k} }}\,{\cos} \left( m\,z \right),
\ee
\be\label{psi2}
\psi_{2_m} \left( z \right) =C_{{m
}}\sqrt {m \left( z+{ k} \right) } \left[ -F(m) \,J_{2}\left( m\left( 
z+{ k} \right) \right) +  N_{{2}}
 \left( m \left( z+{ k} \right)  \right)  \right],
\ee
\ees
where 
\be
F(m)=\frac{F_1(m)}{F_2(m)}=\frac{2\, k\,m\,N_{{1}} \left( m k \right)-3\,N_{{2}}\left( m k \right) }{2\,k\,m\,J_{{1}} \left( m k \right)-3\,J_{{2}} \left( m k \right)  },
\ee  with the normalization factor
\be\label{FN}
  C_m=\frac{1}{\sqrt{\frac{16}{ k\,\pi}+m \left[ F_1(m)^{2}+ F_2(m)^{2} \right]}}.
\ee 
To obtain the above relations we use the Lommel's Formula $J_{{\nu+1}} \left( x \right) N_{{\nu}} \left( x \right) -J_{{\nu}} \left( x
 \right) N_{{\nu+1}} \left( x \right) ={\frac {2}{\pi \,x}}$, and the orthonormality relations of the Bessel functions. From \eqref{psi12} we obtain the probability density
\ben\label{Prob0}
{{|\psi_m \left( 0 \right)|^{2}}}=\frac{16}{\pi^2\,\left|\frac{16}{ k\,\pi}+m \left[ F_1(m)^{2}+ F_2(m)^{2}\right] \right|}.
\een
The probability density, for $m>>{1}/{ k}$ has the predominance of the Bessel function of first kind. On the other hand, for $m<<{1}/{ k}$ the Bessel function of second kind dominates. Then, using the asymptotic forms of Bessel function, $J_{\nu}\left(x \right) \sim\sqrt {{\frac {2}{\pi \,x}}}\,\,{\rm cos}
 \left( -x+1/4\, \left( 2\,\nu+1 \right) \pi  \right)$ and $N_{{\nu}} \left( x \right)\sim\sqrt {{\frac {2}{\pi \,x}}}\,\,{\rm sin}
 \left( -x+1/4\, \left( 2\,\nu+1 \right) \pi  \right)$ for $x>>1$, $J_{{\nu}} \left( x \right) \sim \left( {x}/{2} \right)^{\nu}$ and $ N_{{\nu}} \left( x \right) \sim-{1}/{\pi} \left( 2/x \right)^{\nu}$ for $x<<1$, the probability density assumes the simplified forms:
\ben\label{Prob1}
|\psi_m \left( 0 \right)|^{2}&\sim &{\frac {4\, }{\pi\,(13+4\,{
 k}^{2}{m}^{2})}}
,\quad\,\,\,\quad\,\quad\,\,\,\,\, m>>\frac{1}{ k};\\
 |\psi_m \left( 0 \right)| ^{2}&\sim&\frac83\,\frac{ k
 \,m}{\pi^2}
  ,\quad\,\,\,\quad\,\quad\,\,\,\,\, m<<\frac{1}{ k}.
\een 
The correction to the four-dimensional Newtonian potential
generated by the massive modes, is given by \cite{RS1}
\be\label{03}
V (r) = \frac{M^{-3}_{5}}{r}|\psi_0(0)|^2 + \Delta V (r) ,
\ee 
where the first term is contribution of zero mode and the second term corresponds to
the correction term which is generated by the exchange of KK-modes. Since our problem has no zero mode, then 
\be\label{04}
V(r)=\Delta V (r) = \frac{M^{-3}_{5}}{r}\int^{\infty}_{0}{dm\,{\rm e}^{-m\,r}|\psi_m(0)|^2}.
\ee Our analysis takes place in two distinct regions, where we obtain the probabilities for existence of gravity with continuous mode on the brane at
$z=0$.
Consequently, it is necessary to
divide this integral into two regions
\be\label{07}
\Delta V(r)\sim \frac{M_5^{-3}}{r}\left(
{\frac83\,\frac{ k
 }{\pi^2}}\,\int_0^{\frac{1}{ k}}\, {m\,{\rm e}^{-m\,r} }\,dm\,+4\,\int_{\frac{1}{ k}}^{\infty}\,{\frac {{\rm e}^{-m\,r} }{\pi\,(13+4\,{
 k}^{2}{m}^{2})}}
 \,dm\right)\ee
The Newton law correction for KK-modes is calculated by using equation \eqref{07}
and can be approximated by the form of $\Delta V (r)$ at a distance $r$ given as follows. 

In the limit of the crossover scale $r_c= k$ is very small, i.e., $ k\to0$, the first integral in (\ref{07}) is dominant and gives the following answer
\ben\label{10}
\Delta V(r)\sim{\frac83\,\frac {{M_{{5}}}^{-3}r_c}{\pi^2\,{r}^{3}}},
\een 
which is the same one obtained in the Randall-Sundrum scenario \cite{RS1}.

On the other hand, for the crossover scale being very large, i.e., $r_c=k\to\infty$, the second integral in (\ref{07}) is dominant and
we can get approximately the form of $\Delta V (r)$ at a distance $r$, using the relation ${\rm Ei}(i\,x)={\rm ci}(x)+i\left[1/2\,\pi+{\rm si}(x)\right]$, where ${\rm ci}(x)=-\int_x^{\infty}{\cos\,t/t}$ and ${\rm si}(x)=-\int_x^{\infty}{\sin\,t/t}$, as follows
\ben\label{CiSi}
 V(r)&\sim &\frac{\sqrt{13}}{13}\,\frac{M^{-3}_5}{r_c\,\pi\,r}\,
 \left[ \sin \left( \frac{\sqrt{13}\,r}{2\,r_c} \right) {\rm ci} \left( {
\frac{\sqrt{13}\,r}{2\,r_c}} \right) -\cos \left( \frac{\sqrt{13}\,r}{2\,r_c} \right) {\rm si}
 \left( \frac{\sqrt{13}\,r}{2\,r_c} \right)  \right].
 \label{14}
\een
Let us now discuss the large and small distance behavior comparing with the crossover scale. We use the asymptotic forms: ${\rm ci}(x)\sim\gamma+\ln(x)$ and ${\rm si}(x)\sim x-{\pi}/{2}$ for $x<<1$, $
{\rm ci}(x)\sim{{\rm sin}}(x)/{x}$ and ${\rm si}(x)\sim -{{\rm cos}}(x)/{x}$ for $ x>>1$. 

For small distance, i.e., $r/r_c<<1$,
where $\gamma\approx0.577$ is the Euler-Masceroni constant, we obtain the following form 
\be\label{13}
V \left( r \right) \sim\frac{\sqrt{13}}{13}\,\frac{M^{-3}_5}{r_c\,\pi\,r} \, \left[ \frac{\sqrt{13}\,r}{2\,r_c} \left( \gamma+\ln  \left({
\frac{\sqrt{13}\,r}{2\,r_c}} \right)  \right) -\frac{\sqrt{13}\,r}{2\,r_c}+\frac{\pi}{2} +{\cal O}(r^2) \right], 
\quad r<<{r_c}.
\ee
As we expected, at short distances the potential has the correct $4D$ Newtonian $1/r$
scaling. This is subsequently modified by the logarithmic repulsion term in (\ref{13}).

Finally, for large distance, i.e., $r/r_c>>1$ the potential in Eq.~(\ref{CiSi}) gives
 \be\label{15}
V(r)\sim \frac{2}{13}\,\frac{M_5^{-3}}{\pi^2\,r^2},\quad\quad r>>{r_c}, 
\ee
in accordance with the laws of $5D$ gravity \cite{RG,dvali}.
\\


In summary, it is shown that the gravitational potential becomes
the usual Newton law $(\rightarrow 1/r)$ at short distance and five-dimensional law $(\rightarrow 1/r^2)$ at large distance.
This study showed that from a $5D$ supergravity theory with two scalar fields with standard dynamics the emergence of $4D$ gravity on a BPS asymmetric brane exists even for an asymptotic $5D$ flat space, below a crossover scale,  and the manifestation of extra dimensions does not necessarily occur only at short distances as commonly expected. The complete
behavior is controlled by a crossover scale $r_c$ --- see \cite{023} for similar results and \cite{Neupane:2010ey} for related discussions. This is also in accord with GRS and DGP scenarios, where one has argued that $r_c$ may be related to the present Hubble size.

The authors would like to thanks V.I. Afonso and E. da Hora for discussions, and CAPES,
CNPq, CAPES/PROCAD/PNPD and PRONEX/CNPq/FAPESQ for partial support.


\end{document}